\documentclass[epjST]{svjour}
\usepackage{graphicx}
\usepackage[compress]{natbib} 
\usepackage{times}
\usepackage{color}
\usepackage{slashed}
\usepackage{subfig}
\usepackage{ulem}

\usepackage{amsmath}
\usepackage{amssymb}
\usepackage{hyperref}
\hypersetup{
  colorlinks=true,        
  linkcolor=blue,         
  citecolor=cyan,         
}

\setcitestyle{numbers,square,comma}

\newcommand{\Tr}{\ensuremath{\operatorname{Tr}}}

\newcommand{\ua}{\ensuremath{U(1)_A}}

\def\Fig#1{Fig.~\ref{#1}}

\definecolor{bjcol}{rgb}{0.82,0.1,0.26}
\definecolor{mocol}{rgb}{0.90,0.1,0.70}
\newcommand{\strich}[1]
	{\marginpar{$\left\{ \parbox{2cm}{\\\#1\\[-3mm]} \right.$}}

\newcommand{\mm}{\marginpar{\colorbox{green}{\textbf{BJ}}\\@Mario:}}

\def\roughly#1{\mathrel{\raise.3ex\hbox{$#1$\kern-.75em%
\lower1ex\hbox{$\sim$}}}}

\newcommand{\vev}[1]{\ensuremath{\left\langle #1 \right\rangle}}

\newcommand{\diag}{\ensuremath{\operatorname{diag}}}

\newcommand{\twofigs}{0.49\textwidth}

\graphicspath{
  {./}
  {./figures/}
}

\begin{document}

\title{Nonperturbative quark matter equations of state with vector
  interactions}

\author{Konstantin
  Otto\inst{1}\fnmsep\thanks{\email{konstantin.otto@physik.uni-giessen.de}}
  \and Micaela
  Oertel\inst{2}\fnmsep\thanks{\email{micaela.oertel@obspm.fr}}
    \and Bernd-Jochen Schaefer\inst{1}\fnmsep\inst{3}\fnmsep\thanks{\email{bernd-jochen.schaefer@theo.physik.uni-giessen.de}}}

\institute{Institut f\"{u}r Theoretische Physik,
  Justus-Liebig-Universit\"{a}t Gie\ss en, D-35392 Gie\ss en,
  Germany \and 
LUTH, Observatoire de Paris, PSL Research University,
  CNRS, Universit\'e de Paris, 5 place
  Jules Janssen, 92195 Meudon, France \and
Helmholtz Forschungsakademie Hessen f\"ur FAIR (HFHF),
GSI Helmholtzzentrum f\"ur Schwerionenforschung, Campus Gie\ss en, Germany}


\abstract{Nonperturbative equations of state (EoSs) for two and three
  quark flavors are constructed with the functional renormalization
  group (FRG) within a quark-meson model truncation augmented by
  vector mesons for low temperature and high density. Based on
  previous FRG studies without repulsive vector meson interactions the
  influence of isoscalar vector $\omega$- and $\phi$-mesons on the
  dynamical fluctuations of quarks and (pseudo)scalar mesons is
  investigated. The grand potential as well as vector meson
  condensates are evaluated as a function of quark chemical potential
  and the quark matter EoS in $\beta$-equilibrium is applied to
  neutron star (NS) physics. The tidal deformability and mass-radius
  relations for hybrid stars from combined hadronic and quark matter
  EoSs are compared for different vector couplings. We observe a
  significant impact of the vector mesons on the quark matter EoS such
  that the resulting EoS is sufficiently stiff to support
  two-solar-mass neutron stars.  }

\maketitle

\section{Introduction}
\label{sec:introduction}

With the recent dawn of multimessenger astrophysics new data will
become available with the possibility to further scrutinize models of
the structure of compact objects. In particular, several observational
projects are underway or planned to pin down the compact star equation
of state (EoS) not yet fully known to date. Let us mention for example
precise mass determinations from pulsar timing with
current~\cite{Ozel:2016oaf,Demorest:2010bx,Antoniadis:2013pzd,Cromartie:2019kug}
and future instruments such as the SKA~\cite{Watts:2014tja},
observations of binary neutron star (BNS) mergers by the LIGO/Virgo
collaboration \cite{Abbott:2018fj,GW170817,Abbott:2020uma}
introducing constraints on the EoS via the measurement of the tidal
deformability of the inspiraling stars, and
measurements of the neutron star's x-ray emission giving information
on masses and radii \cite{Watts:2016uzu,Watts:2018iom}, see also the
recent NICER results~\cite{Riley:2019bjaa,Miller:2019bjaa}.
Post-merger oscillations are very sensitive to the dense matter EoS,
too \cite{Bauswein_2019}, and can among others indicate the existence
of a strong first-order phase transition in compact star
matter~\cite{Bauswein:2018bma}, such that future observations by the
LIGO/Virgo/Kagra collaboration or via various projects such as the
Einstein Telescope or the Cosmic Explorer are promising tools to
reduce uncertainties.

However, despite this bright future on the experimental side, a
precise determination of the underlying EoS will not be fully
conclusive because on the theoretical side the EoS alone cannot
resolve for the detailed matter composition and interactions. Only a
full theoretical understanding of dense QCD will enable a satisfactory
final picture which entails the need for a quantitative EoS grounded
in first-principle QCD.

This theoretical challenge is aggravated by the confinement property
of QCD: due to the running of the QCD gauge coupling the fundamental
degrees of freedom lose their significance at low temperatures and
densities and are confined into colorless, composite states, the
hadrons. After confinement, a relatively small residual nuclear
interaction is left which binds the nucleons into atomic nuclei with a
typical binding energy per nucleon of the order of 1-10 MeV. This
energy scale is two orders of magnitude smaller than the confinement
scale but still very strong. Decades of considerable effort together
with constraints from experimental data on nuclei and theoretical
calculations have led to reliable models for the NS crust and
homogeneous nuclear matter up to roughly twice the saturation density,
$\rho_0 \sim 0.16$ fm$^{-3}$, see
e.g. \cite{Oertel:2017fk,Burgio:2018mcr,Tews:2019cap} and references
therein for a discussion. Above this density, not only the models are
less under control, but non-nucleonic degrees of freedom might appear
and the situation becomes more complicated.

Presently, the composition and in particular the phase structure of
the interior of compact stars is not known. In addition to the
possible onset of non-nucleonic hadronic degrees of freedom, such as
hyperons~\cite{Tolos:2020aln,Chatterjee:2015pua,Djapo2008}, mesons,
or $\Delta$-baryons \cite{Kolomeitsev:2016ptu}, quark matter could
exist in the cores of compact objects where due to extreme
gravitational forces densities of up to $\approx 8$-$10 \rho_0$ are
expected \cite{Lattimer:2010uk,Annala:2019puf}. At extremely high
densities $\rho > 60 \rho_0$ perturbative QCD can be used to guide
reliably the construction of the EoS. However, the density of matter
relevant for the description of compact stars is not high enough for a
perturbative treatment to be conducive.  Alternative approaches for a
proper theoretical description of this intermediate density regime
from low-energy QCD are inevitable. Unfortunately, ambitious
first-principle lattice QCD simulations are not applicable in this
regime due to a generic sign problem at finite densities. Therefore,
so far mainly either parameterised interpolations between the
low-density nuclear part and the perturbative QCD regime have been
used, e.g. \cite{Annala:2019puf,Kurkela:2014vha}, generic
parameterisations of the quark matter part, e.g.
\cite{Alford:2017ly,Chamel:2012ea,Alford:2013aca,Zdunik:2012dj} or
phenomenological approaches, such as bag models, Nambu--Jona-Lasinio
(NJL)-type models or the quark-meson coupling model in mean field
approximation have been employed to describe quark matter in this
intermediate density regime,
e.g.~\cite{Benvenuto:1989kc,Torres:2012xv,Baldo:2002ju,Buballa:2003qv,Pereira:2016dfg,Li:2018fk,Zacchi:2019ayh,Zacchi:2015lwa,1506.06846,Zhao:2015rta,Isserstedt:2019pgx,Fischer:2012vc,Peshier:1999ww,Tian:2012zza,Zhao:2015uia,Marczenko:2018uq,Alvarez-Castillo:2016oln}. Some
models include constraints from lattice calculations, see
e.g.~\cite{Schramm:2015hba}, and recently holographic approaches have
been developed, e.g. \cite{Jokela:2018ers,Ishii:2019gta,Jokela:2020piw}. It is
obvious that still much effort is needed and it should in particular
be stressed that the fundamental dynamics of fluctuations is mostly
ignored in mean-field calculations or simple parameterizations, see
e.g.~\cite{Baym2017,Baym:2019iky}.

An alternative approach for the elaboration of the EoS in the
intermediate density regime is based on the functional renormalization
group (FRG) method. It is a promising nonperturbative realization of
Wilsonian renormalization group idea in the continuum and not limited
to small couplings. The straight and controlled computation of the EoS
from first-principle QCD quark and gluonic degrees of freedom becomes
conceivable \cite{Fu:2019hdw}. However, towards lower densities the
quarks cluster into nucleons and the emergence of long-range
correlations between nucleons will increasingly complicate the
QCD-based FRG approach \cite{Weise:2018ukn,Friman:2019ncm,Leonhardt:2019fua}.

Nevertheless, within this framework, the impact of fluctuations in the
(pseudo)scalar interaction channel on the compact star EoS has
recently been studied \cite{Otto:2019zjy}. It was found that the
fluctuations decrease the sound speed of the quark matter even below
the mean-field value of $c_s^2 = 1/3$, leading to a rather soft EoS at
high densities. This rather soft EoS does not allow for the
construction of hybrid stars with a three-flavor quark core in
agreement with present neutron star mass
measurements~\cite{Demorest:2010bx,Antoniadis:2013pzd,Cromartie:2019kug}. Therefore,
here we will extent the work of \cite{Otto:2019zjy} and investigate
the impact of additional vector meson interactions. While the
(pseudo)scalars fields are integrated out within the functional
renormalization group framework, the vector mesons are treated on a
constant background level. This idea has already been employed to
studies of the phase diagram at finite temperature, see
e.g. \cite{Eser:2015bjaa,Drews:2016wpi,Zhang:2017icm,CamaraPereira:2020xla}.

Vector interactions are expected to add repulsion \cite{Song:2019qoh}. This can be seen
from classical models of the nucleon-nucleon
interaction, from phenomenological models allowing
for hyperons in the neutron star core, see
e.g.~\cite{Weissenborn:2011ut,Oertel:2014qza}, or from many
phenomenological quark matter studies, see
e.g. \cite{Alvarez-Castillo:2018pve,Zhu:2018ys,Buballa:1996tm}.  The
inclusion of a repulsive vector interaction in the quark-meson model
should thus stiffen the quark matter EoS, leading to a higher speed of
sound and allowing for constructing hybrid stars compatible with
observations.

Within this first study, we neglect the possibility of diquark pairing
and do not enter the discussion of the extremely rich phase structure
of color superconducting matter in the density range of neutron
stars~\cite{Alford:2007xm,Fukushima:2010bq}. Diquark pairing is
important for transport properties, but, being a Fermi surface
phenomenon, has only little influence on the equation of state we are
focusing on here.

As already mentioned for very low baryon density a neutron star can be
characterized by nonrelativistic nucleons via nuclear forces while at
high densities quarks have more and more of an impact such that the
EoS in the low density regime can be constructed with techniques of
nuclear matter theory. At high baryon density, due to the failure of
first-principle QCD approaches, usually phenomenological quark models
such as NJL-type models are adopted mostly on a mean-field level to
incorporate the dynamics of quarks.  In order to gap the bridge
between these two density regimes nonperturbative effects as the
generation of constituent quark masses by a spontaneous chiral
symmetry breaking must be taken into account.

The work is organized as follows: In the following
Sec.~\ref{sec:vect-degr-freed} a chirally symmetric effective model
with quarks and mesonic degrees of freedom is introduced with
scalar-pseudoscalar as well as axialvector-vector channels to account
for quark matter including chiral symmetry breaking. Later, the focus
lies mainly on the two and three lightest quark flavors relevant for
the strongly interacting intermediate transition regime. The
introduced model setup serves then as truncation for a nonperturbative
integration of the quark and meson dynamics with the FRG in
Sec.~\ref{sec:nonperturbative-eos}. The augmentation of the FRG with
vector mesons is presented in Sec.~\ref{sec:impl-vect-mesons}. Now
equipped with a nonperturbative EoS neutron star properties can be
investigated. The numerical results are presented in
Sec.~\ref{sec:numerical-results} wherein the $\beta$-equilibrated EoS,
mass-radius relation and tidal deformability with and without
strangeness for pure quark matter stars as well as hybrid stars for
various vector couplings are discussed. In Sec.~\ref{sec:summary} a
summary with a conclusion is given. A list of the used input
parameters and some technical details for the numerical solution are
collected in the appendices.

\section{Effective description of  quark matter}
\label{sec:vect-degr-freed}

\subsection{Quark-meson model with vector mesons}
\label{sec:effect-potent-vacu}

A chirally symmetric effective theory is considered with $N_f$ flavors
of constituent quark fields $q$ and $N_f^2$ (pseudo)scalar and
(axial)vector meson fields. With the usual $U(N_f)$ flavor
transformation generators $T_a$ with $a = 0 ,\ldots, N^2_f-1$ the meson
matrices can be rewritten as
\begin{align}
  \label{eq:6}  
\begin{array}{lcll}
  \Phi &:= & T_a (\sigma_a + \mathrm{i} \pi_a )  &
                                                   \qquad\text{for the
                                                   (pseudo)scalar
                                                   mesons and}\\[1ex]
  V_\mu &:= &T_a(\rho_{a,\mu} + \mathrm{i} a_{1 a,\mu}) &
                                                         \qquad\text{for
                                                         the (axial)vector
                                                         mesons. }
\end{array}
\end{align}
Accordingly, the field strength tensor field is given by
$F_{\mu \nu} = \frac{\mathrm{i}}{g_v} \left[ D_{\mu}, D_{\nu}
\right]=\partial_\mu V_\nu - \partial_\nu V_\mu - \mathrm{i} g_v
[V_\mu,V_\nu]$ with the canonical covariant derivative
$D_{\mu} = \partial_{\mu} - \mathrm{i} g_v V_{\mu}$.

For two quark flavors we have the identifications
$\rho_{a,\mu}=(\omega_{\mu}, \pmb{\rho}_{\mu})$ and
$a_{1 a,\mu} =(f_{1,\mu}, \pmb{a}_{1,\mu})$ and for three flavors
$\rho_{a,\mu}=(\omega_{\mu}, \pmb{\rho}_{\mu}, \pmb{K}_{\mu},
\phi_{\mu})$ and
$a_{1 a,\mu} =  ( f_{1,\mu} , \pmb{a}_{1,\mu} , \pmb{K}_{1A, \mu} , \newline
f_{1,\mu} ) $\footnote{The two possible isoscalar axial-vector states
  are $f_{1,\mu} (1285)$ and $f_{1,\mu} (1420)$. }.

The mesons interact via a scalar $g_s$ and a vector $g_v$ Yukawa
coupling with the quarks which is encoded in the quark-meson
Lagrangian in Euclidean space 
\begin{align}
  \begin{split}
    \label{eq:5}
\mathcal{L} &= \bar{q} \left[ \slashed{\partial} + g_s T_a (\sigma_a +
  \mathrm{i} \gamma_5 \pi_a) + g_v T_a \gamma_\mu (\rho_{a,\mu} + \gamma_5 a_{1a,\mu}) \right] q \\
& \quad + \Tr (\partial_\mu \Phi^\dagger \partial_\mu \Phi) + \frac{1}{2} \Tr (F_{\mu \nu} F_{\mu \nu}) + U_\chi (\rho_1, \ldots , \rho_{N_f} )\\
& \quad - c_A [\det \Phi^\dagger + \det \Phi] - \Tr \left[ H (\Phi^\dagger + \Phi) \right] \ .
\end{split}
\end{align}
The anomalous breaking of the axial $\ua$-symmetry is realized via
't Hooft determinants $\xi =[\det \Phi^\dagger + \det \Phi]$ with a
constant parameter $c_A$.  The mesonic (self-)interactions are
parametrized with the chiral symmetric potential
\begin{equation}
\label{eq:4}
U_{\chi} (\rho_1, \ldots , \rho_{N_f} ) \qquad \text{with}\qquad  \rho_n = \Tr [(\Phi^\dagger \Phi)^n]
\end{equation}
and is in general a function of $N_f$ independent chiral invariants
$\rho_n$. The highest chiral invariant $\rho_{N_f}$ is usually omitted
in the chiral potential because it often depends on the other
invariants.

A small explicit breaking of chiral symmetry is implemented with a
linear term in the meson fields $\Tr[H(\Phi^\dagger + \Phi)]$ and a
corresponding symmetry breaking parameter matrix $H$.

We assume isospin symmetric matter such that the two lightest quark
flavors $up$ and $down$ are degenerate in the masses and only one
index $l=u=d$ is needed. Hence, for three quark flavors $N_f=3$ the
explicit symmetry breaking term in \eqref{eq:5} reduces to two
different terms proportional to $-c_l \sigma_l -c_s \sigma_s$ with the
strange quark flavor index $s$. A constant rotation establishes the relation of the 
non-strange-strange basis $(\sigma_l, \sigma_s)$ and the singlet-octet
basis $(\sigma_0, \sigma_8)$ in the scalar meson sector
\begin{equation}
  \label{eq:9}
  \begin{pmatrix}
\sigma_l \\
\sigma_s
\end{pmatrix} = \frac{1}{\sqrt{3}}  \begin{pmatrix}
\sqrt{2} & 1 \\
1 & - \sqrt{2}
\end{pmatrix} \begin{pmatrix}
\sigma_0 \\
\sigma_8
\end{pmatrix}
\end{equation}
such that the isospin-symmetric scalar vacuum condensates are diagonal
in flavor space
\begin{equation}
\label{eq:11}
\vev{\Phi} = T_0 \sigma_0 + T_8 \sigma_8 = \diag_f \left( \sigma_l/2,
  \sigma_l / 2 , \sigma_s / \sqrt{2} \right)\ .
\end{equation}
For $N_f=2$ quark flavors only one independent invariant $\rho_1$ in
the chiral potential survives and hence one explicit symmetry breaking
parameter is needed.

Since the $\ua$-symmetry breaking term scales with the meson fields to
the power $N_f$ the 't Hooft determinant corresponds to a mesonic mass
term for $N_f=2$ and constitutes a mass splitting between the scalar
and pseudoscalar meson multiplets
$\{\sigma_0, \pi_1, \pi_2, \pi_3 \} \leftrightarrow \{\pi_0, \sigma_1,
\sigma_2, \sigma_3 \}$. Hence, the parameter $c_A$ is dropped for
$N_f=2$ and only the first multiplet corresponding to lighter mesons,
i.e., the scalar resonance and the pseudoscalar Goldstone bosons
$\varphi = \left( \sigma, \pmb{\pi}^T \right)^T$, is considered dynamically. In
this way the axial $\ua$-symmetry breaking is assumed to be maximally
broken. For more details see e.g.~\cite{Mitter:2013fxa,Schaefer:2008hk}.

The generalization to finite temperature and baryonic densities is
achieved within the Matsubara formalism, wherein the time-component is
Wick-rotated $t \to -i\tau$ and the imaginary time $\tau$ is
compactified on a circle with radius equal to the inverse temperature
$\beta=1/T$. In general, $N_f$ independent quark chemical potentials
$\mu_f$ can be implemented in the quark part of the Euclidean
Lagrangian, Eq.~\eqref{eq:5}, by
\begin{equation}
\label{eq:12}
\mathcal{L}_{\text{qm}} = \mathcal{L} + 
q^{\dagger} \mu q \qquad \text{with } \mu = \diag_f \left( \mu_u, \mu_d, \ldots, \mu_{N_f} \right)\ .
\end{equation}
However, for cold neutron star matter a weak equilibrium
with neutrinos that leave the star without further interactions is
present such that not all chemical potentials are independent anymore.
For three light quark flavors a common quark chemical potential $\mu$
can be introduced which is related to the baryon number via
$\mu=\mu_B/3$. Respecting the different electrical charges of the
quarks by an additional electron chemical potential $\mu_e$ which is
the negative charge chemical potential one finds
\begin{align}
\begin{split}
\label{eq:13}
\mu_u &= \mu - \frac{2}{3} \mu_e \\
\mu_d &= \mu_s = \mu + \frac{1}{3} \mu_e\ .
\end{split}
\end{align}
Taking electrical charge neutrality of the star into account yields
for the quark and electrical densities
\begin{equation}
\label{eq:14}
\frac{2}{3} n_u - \frac{1}{3} n_d - \frac{1}{3} n_s - n_e = 0 \ ,
\end{equation}
such that only one independent chemical potential remains. Our choice
for two and three quark flavors is the common quark chemical potential
$\mu$. Despite the fact that isospin symmetry is broken in the way the
chemical potentials are introduced we still assume only one light
condensate $\sigma_l$ as approximation. Consequently, even for isospin
asymmetric matter the light quark masses are degenerate
$m_u = m_d = m_l$.

For the following investigation of the equation of state the total
grand potential $\Omega$ is needed. It is given by the logarithm of
the grand partition function,
\begin{equation}
\label{eq:15}
\Omega (T, \mu) = \frac{-T \ln \mathcal{Z}}{V}\ ,
\end{equation}
which in general is a path integral over all involved quantum fields
and hence incorporates all quantum, thermal and density fluctuations
of the studied system. In the literature usually the integration over
the quark fields is performed whereas the dynamics of the remaining
fields are drastically truncated and are taken into account on a
mean-field level, see e.g.~\cite{Zacchi:2019ayh}. In this work we
consider in addition the fluctuations of the remaining mesons via the
FRG method which we will employ in the next Section to evaluate the
grand potential at finite temperatures and densities and determine the
EoS in a nonperturbative manner.

\section{Nonperturbative EoS}
\label{sec:nonperturbative-eos}

\subsection{Functional Renormalization Group Approach}
\label{sec:funct-renorm-group}

For a consistent implementation of successively integrating out
quantum, density and temperature fluctuations from large to small
energy scales we employ Wilson's functional renormalization group idea
\cite{Wilson:1971bg,Wilson:1971dh} in terms of the Wetterich equation
\cite{Wetterich:1992yh}
\begin{equation}
  \label{eq:16}
\partial_t \Gamma_k = \frac{1}{2} \Tr \left[ \partial_t R_k \left(
    \Gamma_k^{(2)} + R_k \right)^{-1} \right]\ .
\end{equation}
This flow equation is a functional differential equation for the
evolution of the scale dependent effective action $\Gamma_k$ where the
logarithmic scale derivative is denoted by
$\partial_t = k \frac{d}{dk}$. The effective average action $\Gamma_k$
interpolates between a microscopic or bare UV action
$S_{\Lambda} = \Gamma_{k \to \Lambda}$ and the full quantum effective
action $\Gamma = \Gamma_{k \to 0}$ in the infrared. It thus governs
the dynamics of the field expectation values after the integration of
quantum fluctuations from the UV scale $\Lambda$ down to the infrared
scale $k_{\text{IR}}$.  The infrared regulator $R_k$ specifies the
regularization of quantum fluctuations near an infrared momentum shell
with momentum $k$. It is a diagonal matrix for mesons and symplectic
for quarks. The scale-dependent IR regulator $R_k$ can be interpreted
as momentum-dependent masses that suppress the infrared modes of the
associated fields. The derivative term $\partial_t R_k$ in
\eqref{eq:16} ensures UV-regularity. The second functional derivative
of the effective average action with respect to the fields of the
given theory is generally denoted as $\Gamma^{(2)}_k$. The trace runs
over all discrete and continuous indices, i.e., color, spinor and the
loop momenta and/or frequencies. Since the full nonperturbative
propagators enters in the flow equation the Wetterich equation is
highly non-linear and includes higher loop contributions in
perturbation theory despite its simple one-loop structure.  The
application of this approach is not restricted to the existence of a
small expansion parameter and hence is applicable in any
nonperturbative regime. For QCD-related applications of this approach
see e.g. the reviews
\cite{Berges:2000ew,Pawlowski:2005xe,Gies2006,Schaefer:2006sr,Braun:2011pp,Dupuis:2020fhh}.

For the explicit solution of the functional equation some truncations
are required that turn it into a system of finite-dimensional partial
differential equations.  Even though any truncation of a functional
equation might induce a certain dependence of physical observables on
the employed regulator the impact can be minimized by choosing
optimized regulators or by implementing RG consistency
\cite{Braun:2018svj}.  Here, a modified three-dimensional flat
regulator has been used \cite{Litim:2001up,Litim:2006ag}.

For the solution of the flow equation \eqref{eq:16} an initial
condition at the UV scale $k=\Lambda$ must be supplemented. A complete
solution of the entire Lagrangian is presently beyond the scope of
this work such that we simplify the system in the following way: The
full dynamical fields in the flow equations are the quarks and the
(pseudo)scalar mesons which affect the effective potential and the
condensates of the remaining vector mesons. For three quark flavors
the UV initial condition for the flow equation \eqref{eq:16} for
$\Gamma_{k=\Lambda}$ reads
\begin{align}
\label{eq:8}
\begin{split}
  \Gamma^{(2+1)}_{k=\Lambda} &= \int d^4x \left\{ \,  \bar{q} \left(\slashed{\partial} +
  g_s T_a (\sigma_a + \mathrm{i} \gamma_5 \pi_a) \right) q \right. \\
& \left. \hspace{1,2cm} + \mathrm{Tr}
\left( \partial_\mu \Phi^\dagger \partial_\mu \Phi \right) +
U^{(2+1)}_{k=\Lambda}(\rho_1, \tilde\rho_2) \right\} \ ,
  \end{split}
\end{align}
wherein the effective potential $U^{(2+1)}_{k=\Lambda}$ depends on the
first two chiral invariants $\rho_1$ and
$\tilde{\rho}_2 := \rho_2 - \rho_1^2/3$, see \eqref{eq:4}. Here
$\rho_2$ has been shifted by $\rho_1^2/3$ for computational
simplicity. For two quark flavors, the effective
potential $U^{(2)}_{k=\Lambda}(\rho_1)$ depends only
  on one chiral invariant. This truncation for the effective action
corresponds to a leading-order derivative expansion with standard
kinetic terms for the (pseudo)scalar meson fields.  In this local
potential approximation (LPA) of the initial action no scalar wave
function renormalizations and no scale-dependence in the Yukawa
couplings between quarks and mesons are taken into account. Note that
in this simple truncation the important dynamical back-reaction of the
mesons on the quark sector of the model is already included. For more
details see the literature, e.g.
\cite{Mitter:2013fxa,Schaefer:2008hk}.

Plugging the truncation \eqref{eq:8} of the action into the Wetterich
equation \eqref{eq:16} yields finally an IR and UV finite flow
equation for the effective potential 
\begin{align}
  \label{eq:frg}
\begin{split}
\frac{\partial U^{(N_f)}_k}{\partial k} &= \frac{k^4}{12 \pi^2} \Bigg
  \lbrace \sum_{b=1}^{2 N_f^2} \frac{1}{E_b} \coth \left( \frac{E_b}{2T} \right) -2
  N_c \sum_f \frac{1}{E_f}\\  
& \hspace{1.5cm} \left[ \tanh \left( \frac{E_f- \mu_f}{2T} \right)
    + \tanh \left( \frac{E_f + \mu_f}{2T} \right) \right] \Bigg\rbrace, 
\end{split}
\end{align}
where the flow of the (pseudo)scalar mesonic degrees of freedom is fully
taken into account. The single-particle energies
$E_{b} = \sqrt{k^2+m_{b}^2}$ include the RG scale dependent
(pseudo)scalar meson masses $m_b$ which are obtained by diagonalizing
the mass entries of the matrix
\begin{equation}
  \label{eq:17}
  M^2_{k,ab} := \frac{\partial^2 U^{(N_f)}_k}{\partial \phi_a
    \partial \phi_b}  \quad \text{for} \quad\phi_a=\left\{ \sigma_a, \pi_a \right\}   \ .
\end{equation} 
Both the potential $U_k^{(N_f)}$ and the mass matrix \eqref{eq:17} are
evaluated at the vacuum expectation value given in Eq. \eqref{eq:11}.
Details and the lengthy explicit expressions of the eigenvalues can be
found in the literature \cite{Mitter:2013fxa,Schaefer:2008hk}.  The
corresponding single-particle energies for the quarks are
$E_u = E_d \equiv E_l = \sqrt{k^2 + g_s^2 \sigma_l^2/4}$ and
$E_s = \sqrt{k^2 + g_s^2 \sigma_s^2/2}$, respectively. The full
thermodynamic potential evaluated at the solution of the gap equation,
i.e., the minimum of the grand potential is obtained by evolving the
system towards the infrared.

As initial UV condition for the flow \eqref{eq:frg} the effective
potential is parameterized for $N_f=2+1$ as
\begin{equation}
  \label{eq:10}
  U^{(2+1)}_{k=\Lambda} (\rho_1, \tilde{\rho}_2)= U^{(2+1)}_{\chi,k=\Lambda} (\rho_1 ,
  \tilde{\rho}_2) - c_A \xi - c_l \sigma_l - c_s \sigma_s
\end{equation}
which contains the scale-dependent chiral effective potential
\begin{equation}
\label{eq:7}
U^{(2+1)}_{\chi, k=\Lambda} (\rho_1 , \tilde{\rho}_2) = a_{10} \rho_1 +
\frac{a_{20}}{2} \rho_1^2 + a_{01} \tilde{\rho}_2
\end{equation}
and the 't Hooft determinants $\xi$ evaluated for the light and strange condensates
\begin{equation}
  \label{eq:19}
\xi = \frac{\sigma_l^2 \sigma_s}{2 \sqrt{2}} \ .
\end{equation}
For three quark flavors only three scale-dependent expansion
coefficients $a_{ij}$ in the chiral potential
$U^{(2+1)}_{\chi,\Lambda}$ are needed and all remaining parameters are
kept constant.  For two quark flavors, the number of coefficients
reduces to two because the second chiral invariant is not independent
anymore.

From the RG point of view at the initial scale it is sufficient to
take only relevant and marginal operators into account since meson
fluctuations are small at high energies (due to their larger masses)
and irrelevant operators are in addition dimensionally suppressed.
However, this does not mean that irrelevant operators can be ignored
in general.  They are generated by the RG flow at smaller scales and
are of relevance, see e.g.~\cite{Pawlowski:2014zaa}.

\subsection{Implementation of vector mesons}
\label{sec:impl-vect-mesons}

The vector mesons are implemented on a mean-field level in the
quark-meson Lagrangian \eqref{eq:5}, i.e., as static background fields
such that their kinetic terms are not of further relevance anymore.
Due to rotational symmetry all components of the vector meson
condensates except the temporal ones are assumed to vanish
\cite{Drews:2014spa}.  For three quark flavors the only non-vanishing
meson fields are in principle the diagonal scalar fields and the
isoscalar vector fields $\omega$ and $\phi$ and the third isovector
vector field $\rho_0^3$. For asymmetric isospin matter on the one hand
two scalar condensates $\sigma_0$ and $\sigma_3$ emerge for two quark
flavors and on the other three scalar condensates $\sigma_i$,
$i=0,3,8$ for three flavors in general. Due to technical
reasons\footnote{Generally for each condensate an extra grid dimension
  for the solution of the corresponding flow equation is necessary
  which is numerically quite time-consuming.} we slightly simplify our
approximation further by dropping the already small isospin breaking
condensate $\sigma_3$ in the flow equation. Hence, to be consistent
within our approximation scheme we also omit the isovector vector
condensate $\rho_0^3$ since this would introduce a further isospin
asymmetry.

For three quark flavors this yields finally a diagonal matrix in
flavor space for the $\omega$- and $\phi$-vacuum expectation values
\begin{equation}
\vev{V_\mu} = \delta_{\mu 0} \frac{1}{2} \diag_f(\omega, \omega, \sqrt{2} \phi)\ .
\end{equation}
This also assumes an ideal quark mixing such that the quark content of
the $\omega$-meson consists purely of \textit{up} and \textit{down}
quarks while the $\phi$-meson is purely \textit{strange}. This in turn
leads to the additional Yukawa-type coupling in the quark sector of
the Lagrangian
\begin{equation}
\mathcal{L}_{\text{vec}} = \frac{g_v}{2} \bar{q} \gamma_0  \diag_f(\omega, \omega, \sqrt{2} \phi) q \ ,
\end{equation}
which can be interpreted as a shift in the corresponding chemical
potentials, giving now rise to modified effective chemical potentials
\begin{align}
\begin{split}
\label{eq:effective_chemical_potentials}
\tilde{\mu}_u &= \mu_u - \frac{g_v}{2} \omega \\
\tilde{\mu}_d &= \mu_d - \frac{g_v}{2} \omega \\
\tilde{\mu}_s &= \mu_s - \frac{g_\phi}{2} \phi
\end{split}
\end{align}
with $g_\phi := \sqrt{2} g_v$.
The constant vector meson vacuum expectation values also contribute to
the mean-field potential
\begin{equation}
\label{eq:vector_potential}
U_{\text{vec}}^{(2+1)} (\omega, \phi)= - \frac{1}{2} \left(m_\omega^2 \omega^2 + m_\phi^2 \phi^2 \right) \ ,
\end{equation}
wherein the negative sign expresses the repulsive nature of the vector
interactions. The mass-like parameters $m_\omega^2$ and $m_\phi^2$ are
basically unconstrained and not to be identified with the physical
vector $\omega$- and $\phi$-meson masses.  They just inherit their
names due to their mass dimension. However, we fix them to the
measured vector meson masses $m_\omega = 782$ MeV and $m_\phi = 1020$
MeV such that the vector meson coupling $g_v$ is the only remaining
free parameter of the system and of the order of one. For two quark
flavors, only the $\omega$-meson is accounted for. However, in the
following we will establish the $N_f=2+1$ flavor equations and
suppress the flavor index.  The two flavor results arise in an obvious
way.

This approximation can now be augmented with the FRG by adding the
vector meson potential $U_{\text{vec}}$ to the scale-dependent chiral
effective potential.  For an arbitrary renormalization scale $k$, the
total effective potential $\tilde{U}_k$ reads
\begin{equation}
\label{eq:full_potential}
\tilde{U}_k = U_k(\rho_1 , \tilde{\rho}_2
, \omega, \phi) +
U_{\text{vec}} (\omega, \phi)\ .
\end{equation}
Since the masses associated with the $\omega$- and $\phi$-bosons are
related to the inverse range of the isoscalar short-distance NN
interactions and are large compared to the relevant low-energy scales
the $\omega$- and $\phi$-fluctuations should be more suppressed. Hence
as a sort of inert degrees of freedom they can be treated as
background fields. Contrarily, the fluctuations in the pseudoscalar
channel (e.g. the sigma and the pions for two flavors) and also the
particle-hole excitations of the quarks around the Fermi surface are
fully taken into account.

The respective condensates are determined in the infrared by solving
the gap equations
\begin{equation}
\label{eq:gap:equations_vector}
\frac{\partial \tilde{U}_{\text{IR}}}{\partial \omega} = 0 = \frac{\partial \tilde{U}_{\text{IR}}}{\partial \phi}
\end{equation}
where $\tilde{U}_{\text{IR}}$ denotes the fully evolved effective
IR potential including all dynamic quark and (pseudo)scalar meson
fluctuations.  The condensates, i.e. the fields at the minimum of the
potential, depend on the temperature and chemical potentials.

The gap equations for the $\omega$- and $\phi$-condensates
\eqref{eq:gap:equations_vector} can be rewritten  
\begin{align}
\begin{split}
\label{eq:gapeq_reformulated}
\omega + \frac{g_v}{2 m_\omega^2} \left.\left(\frac{\partial U_\text{IR}}{\partial \mu_u} + \frac{\partial U_\text{IR}}{\partial \mu_d} \right)\right\vert_{\text{gap}} = 0 =
\phi + \frac{g_\phi}{2 m_\phi^2} \left.\frac{\partial U_\text{IR}}{\partial \mu_s} \right\vert_{\text{gap}} \ .
\end{split}
\end{align}
The subscript "gap" in \eqref{eq:gapeq_reformulated}
labels the gap equation solution meaning that the potential is
evaluated at these field configurations which solve the corresponding
gap equations. Thus, both gap equations for the vector condensates
\eqref{eq:gapeq_reformulated} are self-consistent and can be solved
numerically by root finding.

Finally, the infrared potential $\tilde{U}_{\text{IR}}$ evaluated on the
gap equations can now be identified as the grand potential yielding
the equation of state in a standard thermodynamic manner. For
vanishing temperature the normalized pressure and energy density are given by
\begin{equation}
\label{eq:thermodynamic_relations}
p( \left\{ \mu_f \right\} ) = \Omega (\left\{ 0 \right\}) - \Omega (\left\{ \mu_f \right\})
\qquad \text{and} \qquad \varepsilon = -p + \sum_{f} \mu_f \,
n_f
\end{equation}
with the quark number densities
\begin{equation}
\label{eq:number_densities}
n_f \equiv - \frac{\partial \Omega(T, \left\{ \mu_f \right\})}{\partial \mu_f} = - \left.\frac{\mathrm{d} \tilde{U}_\text{IR}}{\mathrm{d} \mu_f}\right\vert_\text{gap}
\qquad \text{for }f =\left\{ u,d,s
\right\}\ .
\end{equation}
The quark number densities are defined as the derivative of the grand
potential $\Omega (T, \left\{ \mu_f \right\})$ with respect to the
corresponding chemical potentials. Since the implicit dependence of the infrared potential $\tilde{U}_\text{IR}$
on the chemical potentials through the condensates vanishes by virtue of the gap equation, i.e.
\begin{equation}
\label{eq:infrared_potential_chemical_dependence}
\left.\frac{\mathrm{d} \tilde{U}_\text{IR}}{\mathrm{d} \mu_f}\right\vert_\text{gap} = \left. \left( \frac{\partial \tilde{U}_{\text{IR}}}{\partial \mu_f} +
    \frac{\partial \tilde{U}_{\text{IR}}}{\partial \omega}
    \frac{\mathrm{d} \omega}{\mathrm{d} \mu_f} + \ldots
  \right)\right\vert_{\text{gap}} = \left. \frac{\partial \tilde{U}_{\text{IR}}}{\partial \mu_f} \right\vert_{\text{gap}}
  = \left. \frac{\partial U_{\text{IR}}}{\partial \mu_f} \right\vert_{\text{gap}}
\end{equation}
where the ellipses represent similar derivative terms for all other condensates, 
we can identify the derivative terms in \eqref{eq:gapeq_reformulated} with the quark number densities:
\begin{align}
\begin{split}
\label{eq:gapeq_reformulated_densities}
\omega - \frac{g_v}{2 m_\omega^2} (n_u + n_d) = 0 =
\phi - \frac{g_\phi}{2 m_\phi^2} n_s \ .
\end{split}
\end{align}

Note that the gap equation
\eqref{eq:gapeq_reformulated_densities} is solved including the full underlying
nonperturbative contributions from the FRG in the (pseudo)scalar
channel. This is in contrast to a similar two quark flavor FRG study
\cite{Zhang:2017icm} where the gap parameter for the isoscalar
$\omega$-condensate is evaluated from a mean-field flow which ignores
the back-coupling of the FRG flow.

Furthermore, the inclusion of the vector mesons into the
(pseudo)scalar sector appears solely by the replacement of the
chemical potentials in \eqref{eq:frg} with the effective chemical
potentials given in \eqref{eq:effective_chemical_potentials}.

\section{Numerical Results}
\label{sec:numerical-results}

In the following we will present our findings obtained with the FRG
quark-meson truncation including isoscalar vector-mesons for two and
three quark flavors. All results are obtained for $\beta$-equilibrated
and electrical charge neutral matter. Since we are primarily
interested in the physics of older neutron stars for which temperature
effects can be neglected, all flow equations are strictly solved for
vanishing temperature.

\begin{figure}
\centering
\resizebox{0.75\columnwidth}{!}{
\includegraphics{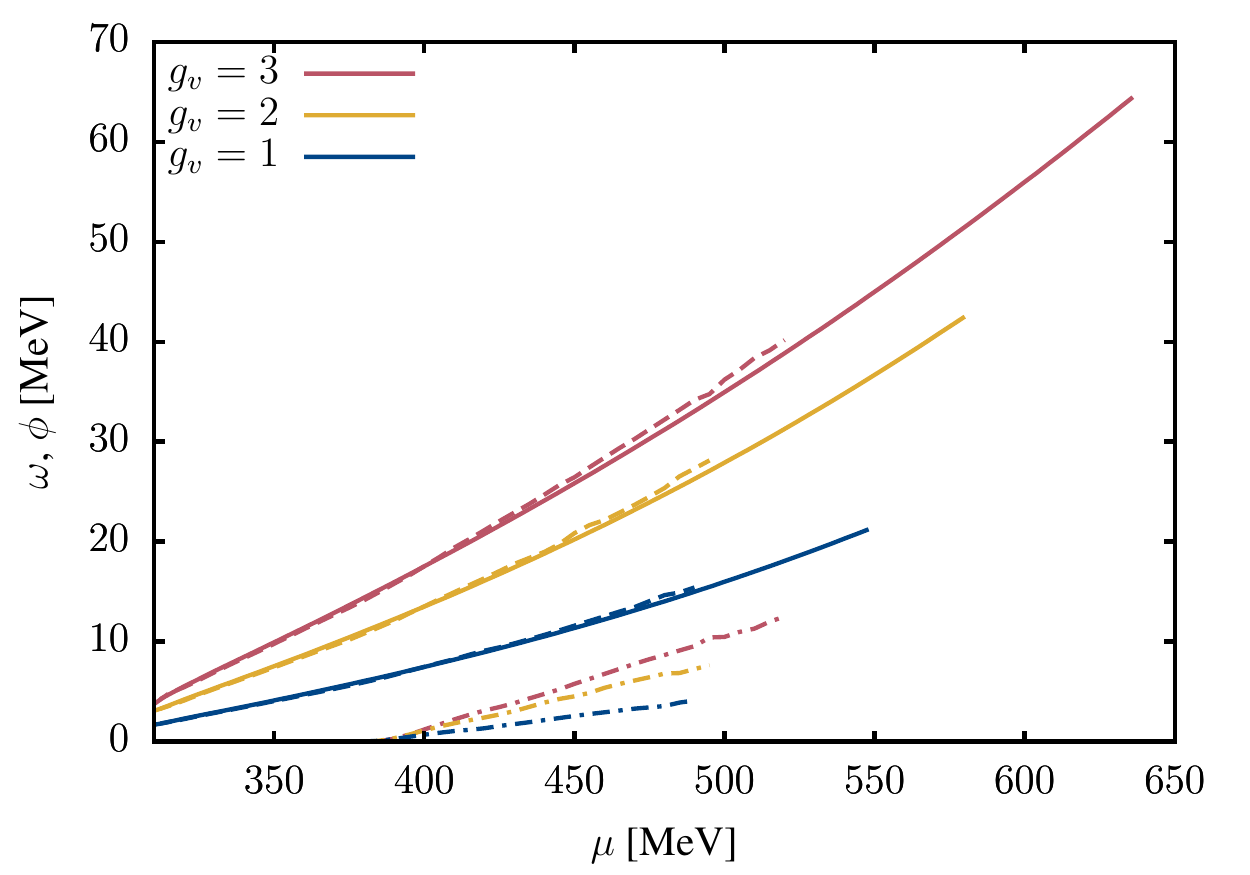}}
\caption{\label{fig:vector_condensates} Vacuum condensates of the
  isoscalar $\omega$-meson (solid line: $N_f=2$, dashed: $N_f=2+1$)
  and $\phi$-meson (dash-dotted) as a function of the quark chemical
  potential. The condensates are evaluated for neutral matter in weak
  equilibrium with vector coupling $g_v$.}
\end{figure}

We have employed two different and complementary numerical solution
strategies for the flow equations as explained in
App.~\ref{sec:numer-solut-deta}. The two quark flavor results are
obtained with an upwind finite difference scheme while for the three
flavor calculations a two-dimensional grid of the two scalar field
variables has been used.  In principle, this enables us to estimate
possible numerical artefacts. We found excellent agreement in
particular at low chemical potentials with no strange quarks
populated, showing the robustness of the numerical scheme. This can be
seen in Fig.~\ref{fig:vector_condensates} where the isoscalar
$\omega$- (dashed and solid lines) and $\phi$-meson (dash-dotted
lines) condensates for three different vector couplings as a function
of the quark chemical potential are shown.  The difference in the
$\omega$-condensate with and without strangeness is negligible. From
the gap equations \eqref{eq:gapeq_reformulated} it is clear that the
vector condensates are proportional to the (respective) number
densities. Hence, for $T=0$ the $\omega$-condensate vanishes in the
chirally broken phase where all occupation numbers are zero, and the
$\phi$-condensate (dash-dotted lines) is zero until $\mu \approx 400$
MeV when the strange quark states get populated. Since an increase in
a vector condensate means a decrease in the respective effective
chemical potential(s),
cf. Eq. \eqref{eq:effective_chemical_potentials}, which in turn
decreases the number density from its initial value, there is always a
unique solution to the vector meson gap equations
\eqref{eq:gapeq_reformulated}. Note that the assumption of one light
chiral condensate for both the \textsl{up} and \textsl{down} flavors
breaks down close to the chiral phase transition under the assumptions
of $\beta$-equilibrium and charge neutrality
\cite{Otto:2019zjy}. Hence, only data points above the chiral phase
transition in the light scalar sector with $\mu > 310$ MeV are
considered.

\begin{figure}
\centering
\resizebox{0.75\columnwidth}{!}{
\includegraphics{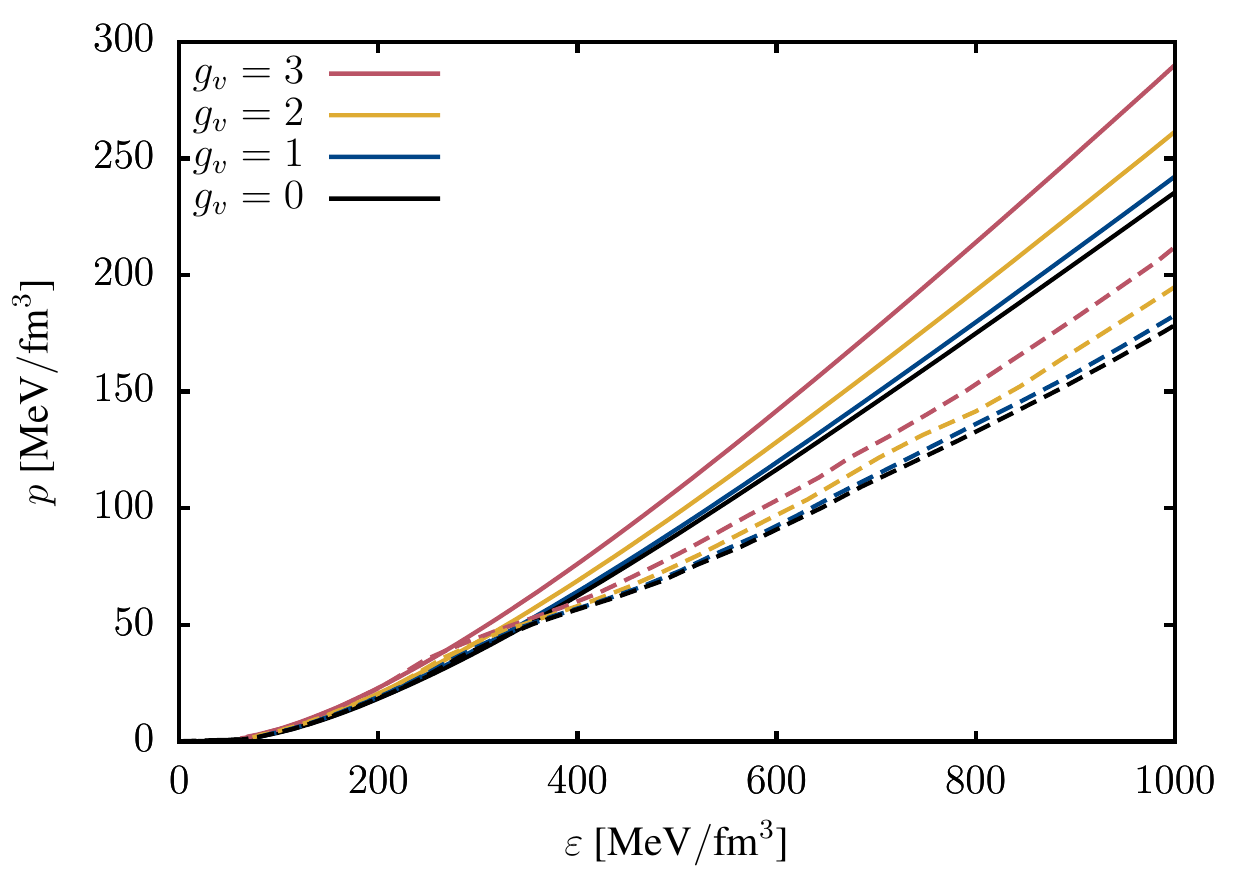}}
\caption{\label{fig:eos_vecmes} Equation of state of the FRG
  quark-meson model with vector mesons and coupling $g_v$. Weak
  equilibrium and charge neutrality conditions have been
  imposed. Solid lines correspond to $N_f=2$ quark matter and dashed
  lines to $N_f=2+1$.}
\end{figure}

The effect of the background isoscalar vector mesons on the equation
of state is displayed in Fig. \ref{fig:eos_vecmes}. In this figure the
(normalized) pressure obtained with the FRG quark-meson truncation
including the vector meson condensates is displayed for different
vector couplings $g_v$ as a function of the corresponding energy
density. As already mentioned, $\beta$-equilibrium and charge
neutrality have been implemented. The two quark flavor EoS (solid
lines) is stiffer than the corresponding EoS with strangeness (dashed
lines) for energy densities beyond the onset of strangeness. This is
not astonishing since an additional degree of freedom generally
reduces pressure and thus softens the EoS.

Vector mesons contribute to the EoS with two effects. Firstly, since
the vector meson potential \eqref{eq:vector_potential} gives a
negative contribution to the grand potential, it follows from
\eqref{eq:thermodynamic_relations} that an increasing vector meson
condensate leads to an increasing overall pressure. Secondly, at the
same time, the effective chemical potentials are lowered which reduces
the contributions to the pressure and also the energy density via the
particle densities from the (pseudo)scalar and quark
sectors. Altogether, a larger vector coupling leads to an increase of
the EoS's stiffness both for $N_f=2$ and $N_f=2+1$ quark flavors as
expected.

\begin{figure}[htb!]
  \subfloat[\label{fig:tov_M_R_2} $N_f=2$]{\includegraphics[width =
    \twofigs]{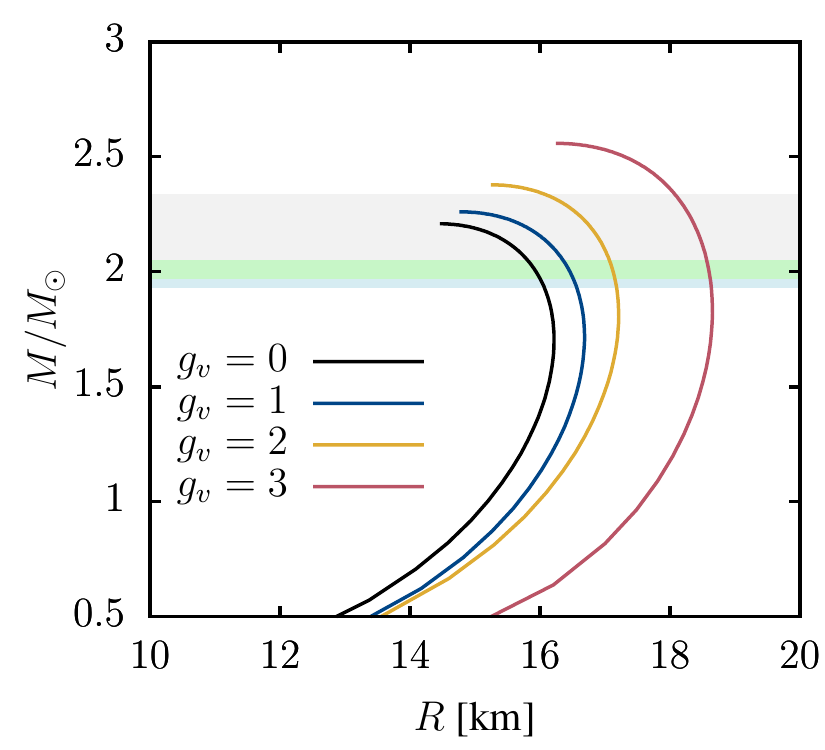}} 
\subfloat[\label{fig:tov_M_R_2p1} $N_f=2+1$]{\includegraphics[width =
    \twofigs]{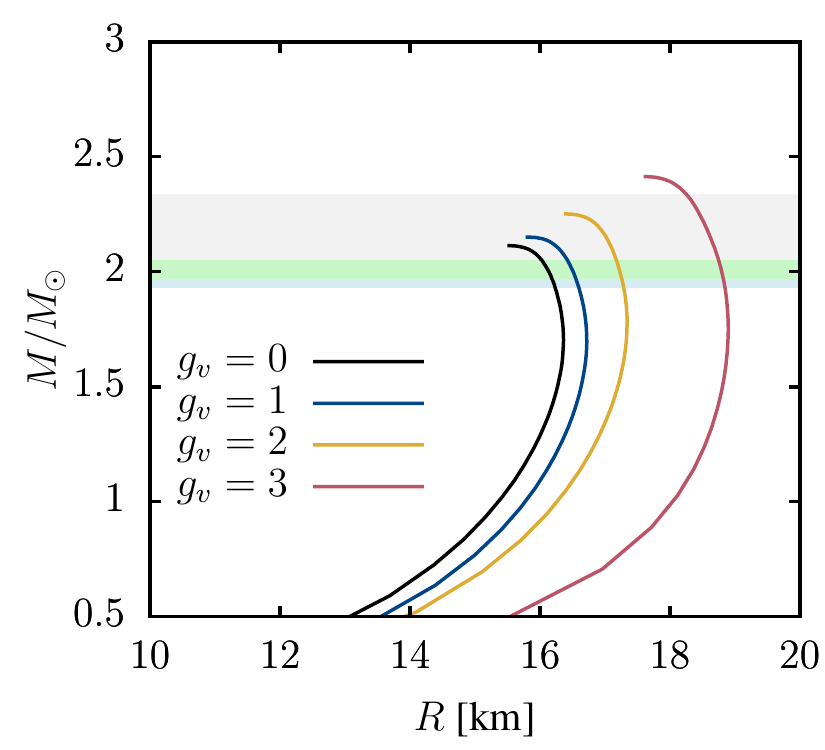}}
  \caption{\label{fig:tov_M_R} Mass-radius relations for pure quark
    matter stars with various vector couplings $g_v$ based on the EoS
    in Fig.~\ref{fig:eos_vecmes}. Horizontal bands: PSR J1614-2230
    (blue), PSR J0348+0432 (green) and MSP J0740+6620 (gray)
    mass measurements
    \cite{Demorest:2010bx,Antoniadis:2013pzd,Cromartie:2019kug}.}
\end{figure}

Knowing now the EoS for pure quark matter, the mass-radius relation of
a compact star can be obtained as solution of the
Tolman-Oppenheimer-Volkoff equations, assuming a perfect fluid and a
non-rotating star. Such pure quark stars \cite{Haensel_86} could exist
under the hypothesis of absolutely stable strange quark
matter~\cite{witten84,Farhi:1984qu}. Within our setup, quark matter is
not absolutely stable, but it is nevertheless instructive to
investigate the mass-radius relation of quark stars with different
strengths of the vector coupling.  The results are summarized in
Fig.~\ref{fig:tov_M_R}, in the left panel for two flavor quark matter
and in the right panel including strangeness.  The colored horizontal
bands indicate the measured two-solar-mass
pulsars~\cite{Demorest:2010bx,Antoniadis:2013pzd,Cromartie:2019kug}.
Increasing the vector coupling shifts the masses to larger values and
increases the radii. Maximum masses are all compatible with observed
pulsar masses, and radii are generally larger than current neutron
star observations
suggest~\cite{Ozel:2016oaf,Abbott:2018fj,Riley:2019bjaa,Miller:2019bjaa},
see also \cite{Otto:2019zjy}.

\begin{figure}[htb!]
	\centering
  \subfloat[\label{fig:eos_hybrid_vecmes_2} $N_f=2$]{\includegraphics[width =
    \twofigs]{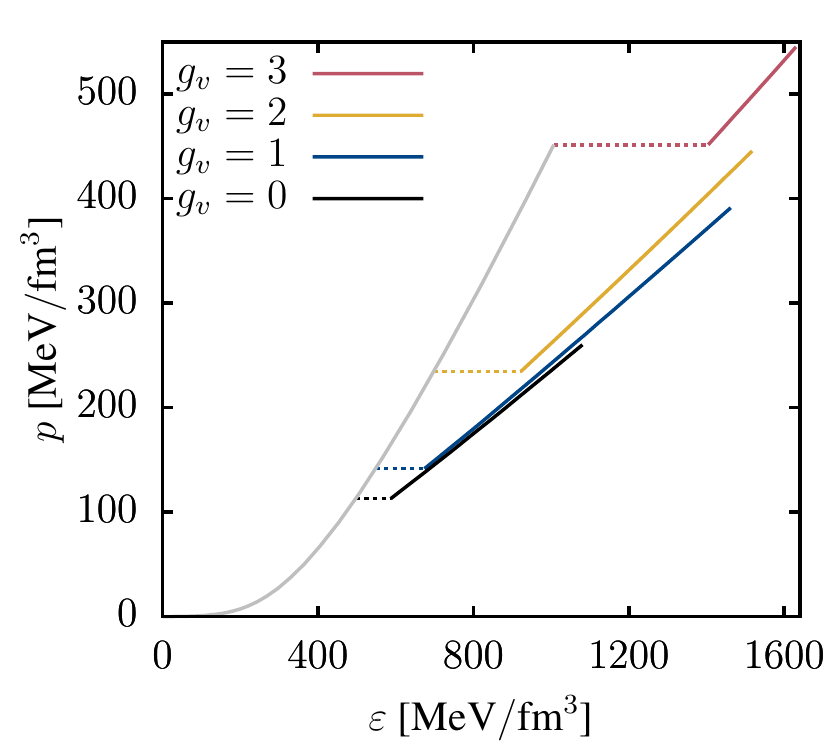}} 
\subfloat[\label{fig:eos_hybrid_vecmes_2p1} $N_f=2+1$]{\includegraphics[width =
    0.396\textwidth]{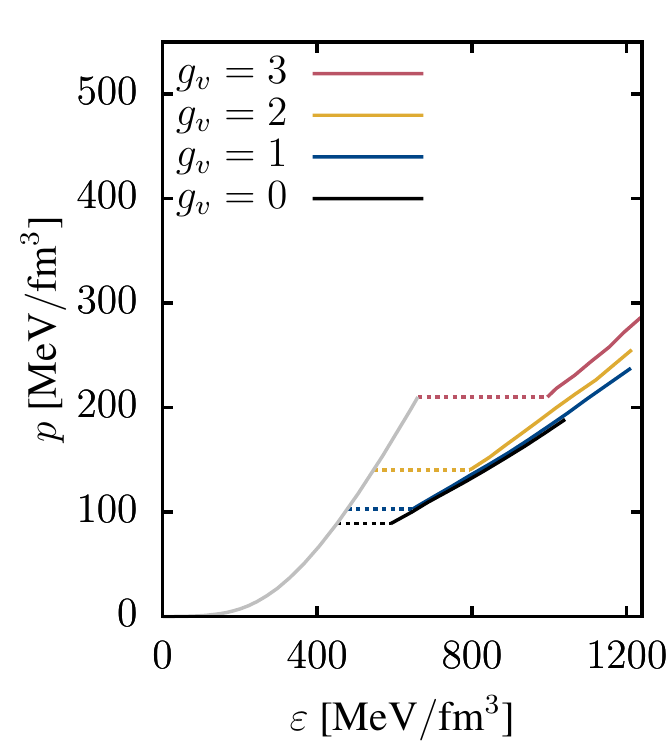}}
  \caption{\label{fig:eos_hybrid_vecmes} Equation of state for hybrid
    matter. The nucleonic phase is described by the HS(DD2) EoS (gray
    color) and the quark matter phase by the FRG quark-meson EoS with
    vector mesons (see Fig. \ref{fig:eos_vecmes}). Both phases
    separately meet weak equilibrium and charge neutrality and are
    connected via a Maxwell construction. For $N_f=2+1$, energy
    densities larger than $\varepsilon \sim 1200 \, \text{MeV/fm}^3$
    corrresponding to $\mu >
    500 \, \text{MeV}$ are dropped.}
\end{figure}

As mentioned above, within our setup quark matter is not absolutely
stable and we will now turn to the construction of a hybrid matter
EoS.  Note that while generally an increasing vector coupling
increases the EoS's stiffness the pressure decreases with increasing
vector coupling for a given quark chemical potential.  This can be
explained by the aforementioned reduction of the effective chemical
potential in the quark loop, leading to an overall pressure reduction
for a given chemical potential. This has significant consequences for
the possible occurrence of hybrid stars, i.e. neutron stars with a
quark matter core. We construct a hybrid EoS by assuming two separate
phases, a nucleonic phase described by the HS(DD2) hadronic equation
of state~\cite{Typel:2009sy,Hempel:2009mc} and a quark matter phase
described by the present FRG quark-meson EoS with additional vector
meson interactions. The results are shown in
\Fig{fig:eos_hybrid_vecmes}, where the pressure as a function of the
energy density is displayed (left panel for two quark flavors and
right panel for $N_f=2+1$). Both phases are separated by a clear
boundary and individually fulfill the weak equilibrium and charge
neutrality conditions. A first-order transition is obtained via a
Maxwell construction (horizontal dotted lines in the figure).  It can
be characterized by an onset energy density $\varepsilon_{\rm trans}$
in the hadronic phase and a gap $\Delta \varepsilon$ given by the
difference between the energy density in the quark phase at the end of
the transition and the onset $\varepsilon_{\rm trans}$. The onset
energy density also defines the transition pressure
$p_{\rm HS(DD2),trans} \equiv p_{\rm HS(DD2)}(\varepsilon_{\rm
  trans})$.

Due to the decreasing quark matter pressure at increasing vector
interaction strength, the phase transition gradually moves to higher
quark chemical potentials, i.e. to a higher intersection pressure and
a higher $\varepsilon_{\rm trans}$. $\Delta \varepsilon$ also increases
for increasing $g_v$.  For $N_f=2+1$, the transition generally occurs
at lower pressures and with larger $\Delta\varepsilon$ than for
$N_f=2$ due to the additional strange degree of freedom in the quark
matter EoS.

The size of the discontinuity of the energy density
$\Delta\varepsilon$ determines, too, the stability of the hybrid star
against gravitational collapse: a large discontinuity destabilizes the
star immediately at the transition point $p = p_{\rm trans}$ whereas
for a small discontinuity a small quark core forms and the star
remains stable. This scenario can be summarized in terms of the Seidov
limit \cite{Seidov:1971aa}
\begin{equation}
\label{eq:1}
\frac{\Delta\varepsilon_{\rm crit}}{\varepsilon_{\rm trans}} =
\frac{1}{2} + \frac{3}{2}\frac{p_{\rm trans}}{\varepsilon_{\rm
    trans}}\ .
\end{equation}
$\Delta \varepsilon_{\rm crit}$ denotes here the threshold value below
which a stable hybrid star branch is connected to a hadronic star
branch. Thus, above the Seidov limit the sequence of stars become
unstable immediately. All EoS displayed in
Fig.~\ref{fig:eos_hybrid_vecmes} remain well below this limit.

\begin{figure}
\centering
\resizebox{0.75\columnwidth}{!}{
 \includegraphics{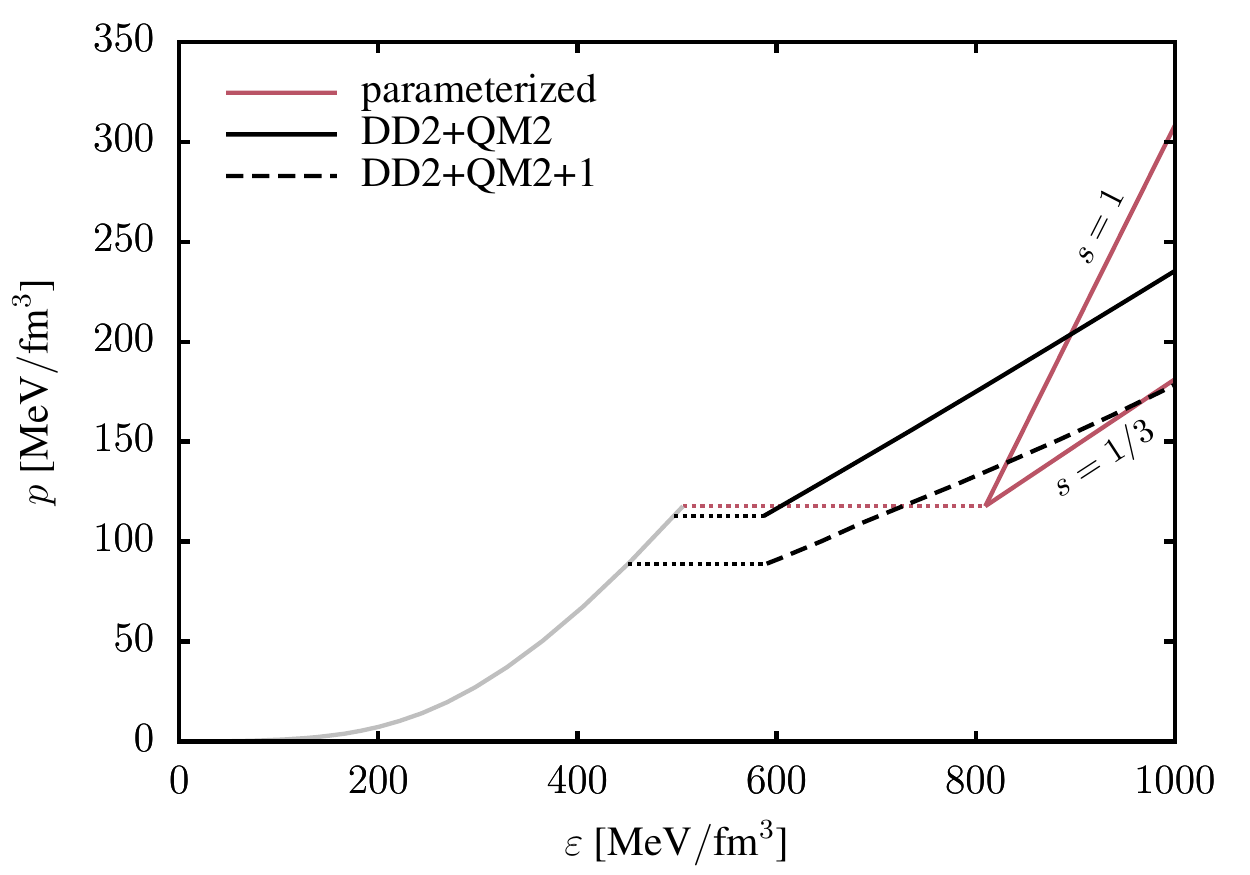}}
\caption{\label{fig:eos_hybrid_sym} Similar to
  \Fig{fig:eos_hybrid_vecmes}: composite EoS for the FRG quark matter
  (QM2 and QM2+1) and HS(DD2) nuclear model (gray color) compared with the hadron
  quark EoS QHC19~\cite{Baym:2019iky}.  A combination of the HS(DD2)
  EoS with a parameterized quark matter EoS~\cite{Alford:2017ly} for
  two different speed of sound values $c_s^2 \equiv s = 1$ and $1/3$ are also
  shown. }
\end{figure}

If $\Delta\varepsilon > \Delta\varepsilon_{\rm crit}$, a so-called
``third family''~\cite{Schertler:2000xq} stable sequence of hybrid
stars may exist at higher central densities for certain conditions,
leading eventually to twin or even triplet configurations
\cite{Benic:2014jia,Alford:2017ly}.  The conditions for the existence
of such twin or triplet configurations has been discussed in detail in
\cite{Alford:2017ly}, characterising the transition by the two
parameters $\varepsilon_{\rm trans}$ and $\Delta\varepsilon$ together
with a constant speed of sound parameterisation of the quark phase. In
Fig.~\ref{fig:eos_hybrid_sym} we show, together with the FRG hybrid
EoS with $g_v = 0$, such a parameterisation
\begin{equation}
  p(\varepsilon) = 
  \left\{
    \begin{array}{lrl}
            p_{\rm HS(DD2)} (\varepsilon)& , &  \varepsilon <
                                               \varepsilon_{\rm trans}  \\
            p_{\rm HS(DD2)} (\varepsilon_{\rm trans})& , &
                                                           \varepsilon_{\rm trans}<\varepsilon < \varepsilon_{\rm trans}
                                                           + \Delta \varepsilon \\
            p_{\rm HS(DD2)} (\varepsilon_{\rm trans})
            + s \ [\varepsilon -  (\varepsilon_{\rm trans} + \Delta
            \varepsilon) ]& , & \varepsilon >
                                \varepsilon_{\rm trans} + \Delta \varepsilon \ .
    \end{array}
  \right.
\end{equation}
$\varepsilon_{\rm trans}$ has thereby been chosen very close to the
transition density in the FRG two-flavor hybrid EoS, whereas
$\Delta\varepsilon/ \varepsilon_{\rm trans} = 0.6$ is close to the
Seidov limit.  It is obvious, as already anticipated from the Seidov
limit, that the FRG hybrid stars remain stable. Two different values
for the speed of sound in the quark phase have been chosen. The first
one, $s = 1/3$, is close to the FRG sound speed, whereas the second
one $s = 1$ represents the causality limit, i.e. the stiffest possible
quark matter EoS. As already discussed in \cite{Otto:2019zjy}, for
$s=1/3$, the pressure in the quark matter phase is not sufficient to
counteract the strong gravitational pull due to the large energy
density of the quark core, cf. Ref. \cite{Alford:2017ly}, and thus
does not support a stable hybrid star branch. For $s=1$, a stable
third family branch with twin configurations
exists~\cite{Otto:2019zjy}. Since even with the inclusion of vector
interactions $\Delta\varepsilon$ does not exceed the Seidov limit and
the sound speed is only insufficiently increased in the quark phase,
the findings discussed here for the case $g_v = 0$ remain valid for
nonzero vector coupling. We thus confirm the conclusion of
\cite{Otto:2019zjy} that the occurrence of twin stars in our model is
ruled out due to the small energy gap at the phase transition from
nuclear matter to quark matter and due to the small stiffness of the
quark matter EoS.

\begin{figure}[htb!]
  \subfloat[\label{fig:tov_M_R_hybrid_2} $N_f=2$]{\includegraphics[width =
    \twofigs]{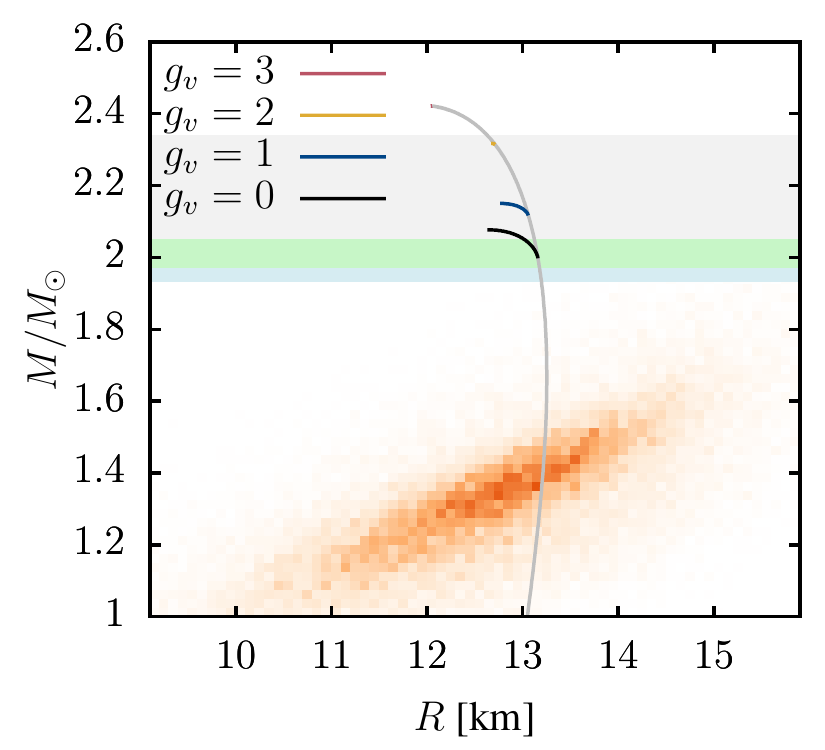}} 
\subfloat[\label{fig:tov_M_R_hybrid_2p1} $N_f=2+1$]{\includegraphics[width =
    \twofigs]{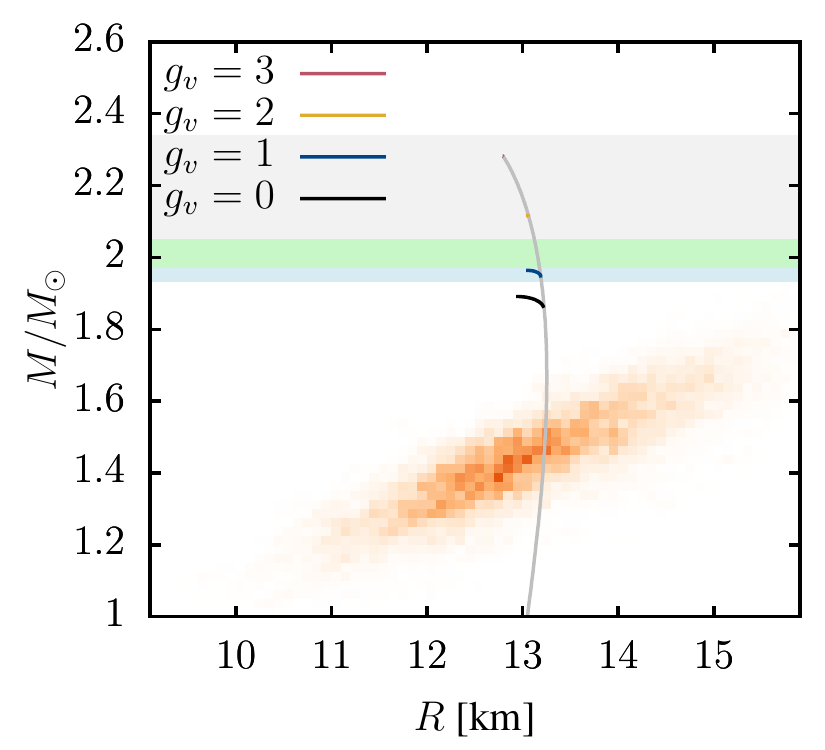}}
  \caption{\label{fig:tov_M_R_hybrid} Mass-radius relations for hybrid
    stars from the combined HS(DD2) and quark-matter EoSs (see
    Fig. \ref{fig:eos_hybrid_vecmes}) for various vector couplings
    $g_v$. Purely nucleonic stars governed only by the HS(DD2) EoS are
    depicted in gray color. The horizontal bands are similar to
    Fig. \ref{fig:tov_M_R}.  Additionally, the posterior probability
    distributions for the mass-radius correlations from the two
    independent recent NICER analyses are depicted (left panel: Riley
    et al. \cite{Riley:2019bjaa}, right panel: Miller et
    al. \cite{Miller:2019bjaa}).}
\end{figure}

The shift of the phase transition in a hybrid star to higher densities
with increasing vector coupling can also be seen in the mass-radius
relations, shown in \Fig{fig:tov_M_R_hybrid} from the combined HS(DD2)
and the present FRG quark-matter EoSs (left panel two quark flavors
and right panel with strangeness).  An increase of the vector coupling
leads to a continuously smaller quark matter core in the hybrid star,
but an increasing maximum mass. Especially for $N_f=2+1$ (right
panel), where without vector interactions the two solar mass limit
cannot be satisfied, the maximum mass is in agreement with current
observations for $g_v \gtrsim 1$. However, a quark matter core is only
found as a small, continuous branch.  For example, for $g_v=1$ the
$N_f=2+1$ hybrid star model yields for the heaviest stable star a
quark matter core with radius around $3.2$ km, constituting about
$4\%$ of the star's total gravitational mass. For $g_v=2$, the
heaviest star's quark core radius is $1.6$ km and makes up only
$0.6 \%$ of its total mass.  Concerning the possibility of twin stars,
the mass-radius relations confirm the absence of twin configurations
within our model.

Below the onset of quark matter, the mass-radius relation coincides
with the nuclear HS(DD2) one as it should. This means that the
properties of stars with masses below 1.8 $M_\odot$ are given entirely
by the HS(DD2) EoS. Among others, the HS(DD2) EoS leads to a
relatively large radius for intermediate mass stars which seems,
although being in agreement with recent NICER results, to be
disfavored by some radius determinations, see
e.g.~\cite{Ozel:2016oaf}. We expect that a hybrid construction with a
nuclear EoS which features a smaller radius for intermediate mass
stars, does not result in any quark core in the stars. Since with
increasing vector coupling, the pressure of the quark matter EoS as a
function of the chemical potential is decreased, shifting thus the
transition, as noted already for the HS(DD2) EoS, to higher densities
and above the central densities of stable neutron stars for most
nuclear EoSs.

\begin{figure}[htb!]
  \subfloat[\label{fig:tidal_2_combined} $N_f=2$]{\includegraphics[width =
    \twofigs]{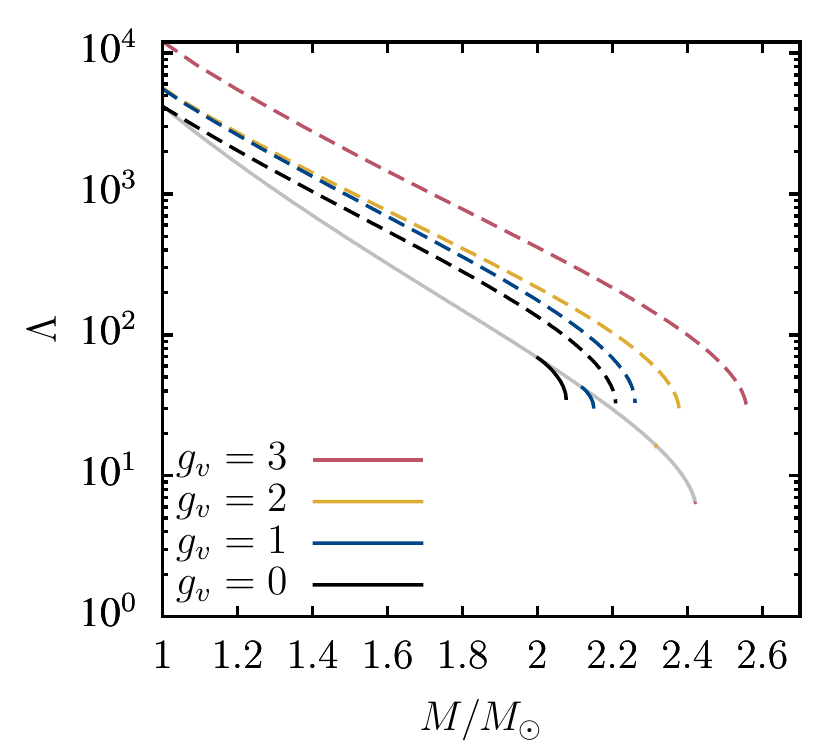}} 
\subfloat[\label{fig:tidal_2p1_combined} $N_f=2+1$]{\includegraphics[width =
    \twofigs]{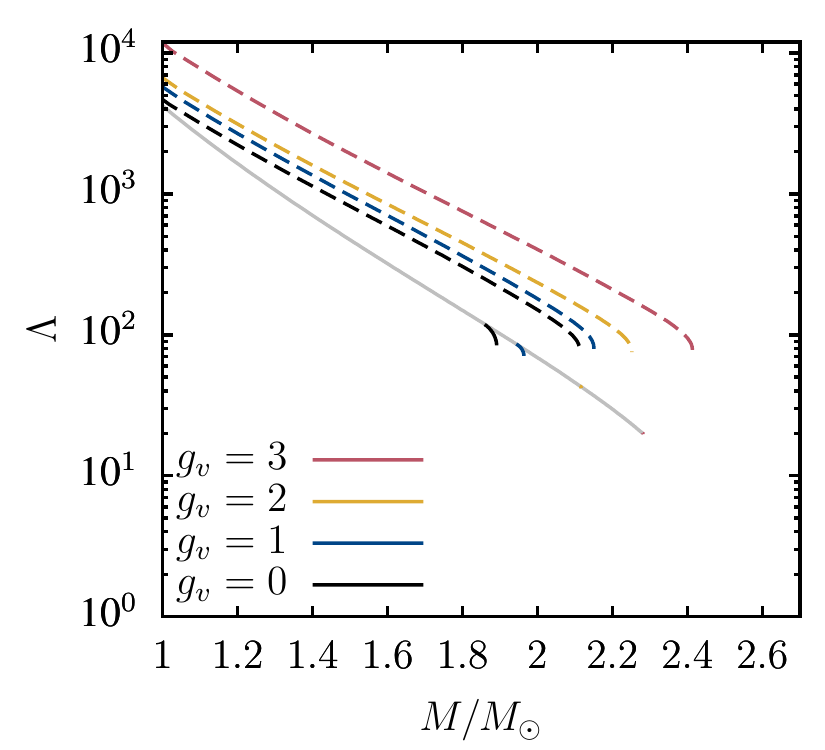}}
  \caption{\label{fig:tidal_combined} Dimensionless tidal
    deformabilities of pure quark (dashed lines) and hybrid stars
    (solid lines) with different vector couplings $g_v$ as a function
    of gravitational mass. Similar to Fig.~\ref{fig:tov_M_R_hybrid},
    the deformabilities of purely nucleonic stars are shown in gray.}
\end{figure}

Another interesting quantity that is experimentally accessible is the
tidal deformability of neutron stars.  For a static, spherically
symmetric star, placed in a static external quadrupolar tidal field
$\mathcal{E}_{ij}$, the tidal deformability $\lambda$ can be defined
to linear order as
\begin{equation}
Q_{ij} = -\lambda \mathcal{E}_{ij}\ ,
\end{equation}
where $Q_{ij}$ represents the star's induced quadrupole moment. The
tidal parameter $\lambda$ can then be computed from a perturbation of
the spherical TOV solution, see \cite{Hinderer:2009ca} for more
details. The results for the dimensionless tidal deformability
$\Lambda = \lambda / M^5$ as a function of the star's gravitational
mass are shown in Fig. \ref{fig:tidal_combined} for hybrid and quark
stars with two and three quark flavors, respectively, and different
values of the vector coupling.

The pure quark stars lead to tidal deformabilities which are
significantly too large compared with the GW170817
observations~\cite{GW170817,Abbott:2018fj}. This is also in line with
findings from parameterized EoS's in \cite{Zhao:2018nyf}. As discussed
before, we do not expect pure quark stars to exist within our setup
since quark matter is not absolutely stable. The hybrid star tidal
deformabilities only differ from the HS(DD2) ones close to their
respective maximum masses, i.e., the quark cores are too small to have
an impact on the tidal deformabilities for all inspiral stars of
masses below $\approx 1.8 \, M_\odot$ or even higher, depending on
$g_v$. The HS(DD2) tidal deformability is in slight tension with the
GW170817 observations. A hybrid construction with other nucleonic EoS
leading to lower tidal deformabilities could thus be
appropriate. However, in \cite{Otto:2019zjy} we found no intersection
of the quark matter EoS with other such nucleonic EoS, since the
nucleonic pressure over the entire relevant range exceeded the quark
matter one for given chemical potentials. Since with increasing vector
coupling, the pressure is reduced for given chemical potential, we
confirm that we do not find hybrid stars with lower tidal
deformabilities within the present FRG approach to the quark matter
EoS.

\section{Summary and conclusions}
\label{sec:summary}

Based on a previous work \cite{Otto:2019zjy} a two- and three flavor
quark-meson model that fully incorporates chiral symmetry breaking has
been augmented with vector mesons to investigate the quark matter EoS
relevant for neutron star physics. Quantum and density fluctuations of
the quarks and the (pseudo)scalar meson channels beyond the mean-field
approximations are treated using the nonperturbative functional
renormalization group method wherein the coupling of the isoscalar
vector mesons to the FRG flow is taken into account.

As an application pure quark stars as well as hybrid stars are
analyzed for different vector meson couplings always in
$\beta$-equilibrated neutral matter. Since quark matter within our
setup is not absolutely stable we construct hybrid stars where the EoS
obtained with the FRG are combined with a nuclear EoS via a Maxwell
construction. In general, an increase of the vector interaction
increases the EoS's stiffness while the pressure decreases for a given
chemical potential due to a reduction of the effective chemical
potential in the quark loop.

Similar to the previous findings of the FRG EoS without vector
interactions the inclusion of strangeness softens the EoS at high
densities. This is in contrast to the addition of repulsive vector
interactions to the system which stiffens the EoS. However, because of
the decreasing quark matter pressure at increasing vector interaction
strength the phase transition is shifted to higher quark chemical
potentials. Also the energy density gap at the transition increases
with increasing vector couplings. Including strangeness, the
transition generally occurs at lower pressures but with larger energy
density gaps which is plausible due to additional strange degree of
freedom in the EoS.

As a consequence pure quark matter as well as hybrid stars with vector
interactions lead to larger maximum masses such that the stars are
consistent with existing observed pulsar masses. For pure quark stars,
the radii are generally larger than observations suggest. Note that
for hybrid stars without vector interactions the two solar mass limit
cannot be reached when strangeness is included.

However, even though the hybrid stars' maximum mass increases with
vector interaction, the quark matter core becomes smaller. Despite the
fact that a quark matter core is only found as a small continuous
branch in the mass-radius relation the largest core radius is found
for a vanishing vector coupling. Within our setup the possibility of
twin stars can be excluded.

Furthermore, hybrid stars with lower tidal deformabilities can also be
excluded with the present FRG setup. The tidal deformability is mainly
determined by the nucleonic EoS and differs from it only in the
vicinity of its maximum mass where a quark matter core forms. The
nucleonic pressure required to produce smaller tidal deformabilities
would be too large to allow for a hybrid construction over the
relevant range of chemical potentials.

In total, regarding repulsive vector interactions the occurrence of
hybrid stars with extensive quark cores seems to be disfavored within
our setup, in particular in view of experimental restrictions on the
masses, radii, and tidal deformabilities. Besides these findings, a
number of open issues remain: A systematic improvement of the employed
FRG truncation, specifically including the running of higher
derivative couplings in the (pseudo)scalar channel, might lead to
further insights on the impact of fluctuations and on the robustness
of the approach. Finally, the role of many-body quark correlations
relevant for a more realistic description of dense matter -- such as
residual six-quark or diquark correlations in the quark phase -- needs
to be considered in the future.

\subsection*{Acknowledgments}
This work was supported in part by the Helmholtz International Center
for FAIR within the LOEWE initiative of the State of Hesse, the
Deutsche Forschungsgemeinschaft (DFG) through the grant CRC-TR 211
“Strong-interaction matter under extreme conditions” with project
number 315477589 - TRR 211, and the BMBF under contract
No. 05P18RGFCA.
K.O. acknowledges funding by the German Academic Scholarship
Foundation and the Helmholtz Graduate School for Hadron and Ion
Research (HGS-HIRe) for FAIR.  M.O. acknowledges financial support
from the action ``Physique fondamentale et ondes gravitationnelles''
of Paris Observatory.

\appendix

\section{Input Parameters}
\label{app:inputparam}

In this appendix all input parameters for the FRG calculations are
given. For further information see also
\cite{Mitter:2013fxa,Schaefer:2008hk,Otto:2019zjy}.
We start with $N_f=2+1$ quark flavors and then simplify to $N_f=2$
quark flavor.  The finite pseudoscalar masses of the pions,
$m_\pi = 138$ MeV, and kaons, $m_K = 496$ MeV, are fixed via the
explicit chiral symmetry breaking terms
$c_l = (120.73 \, \text{MeV})^3$ and $c_s = (336.41 \, \text{MeV})^3$.
Without an explicit symmetry breaking all pseudoscalar masses would
vanish in a chirally invariant or spontaneously broken theory due to
the Goldstone theorem. The summed squares of the $\eta$ and $\eta'$
masses $m_\eta^2+m_{\eta'}^2 = (1103.2 \, \text{MeV})^2$ are
reproduced with the axial $U(1)_A$ symmetry breaking parameter
$c_A = 4807.84$ MeV. The remaining three parameters in the ultraviolet
effective potential are fixed with the (broad) sigma meson resonance
mass which we have chosen to be of the order of $m_\sigma = 560$ MeV
and the two vacuum condensates $\sigma_{l,0} = 92.4$ MeV and
$\sigma_{s,0} = 94.5$ MeV that yield the pion and
  kaon decay constants, $f_\pi=92.4$ MeV and $f_K=113$ MeV.  The
constituent light and strange quark masses follow from a single Yukawa
coupling $g=6.5$, i.e.  $m_l = g \sigma_{l,0}/2 \approx 300$ MeV and
$m_s = g \sigma_{s,0}/\sqrt{2} \approx 434$ MeV.  For $N_f=2$, we
proceed in an analogous fashion. Only the remaining non-strange
condensate $\sigma_{l,0}$ and $m_\sigma$ are set by the two free
parameters in the chiral potential in the ultraviolet whereas $m_l$
and $m_\pi$ are fixed by $g$ and $c_l$ as before.

The used input parameters of the chiral potential for an initial UV
cutoff of $\Lambda = 1$ GeV and an infrared cutoff of
$k_\text{IR} = 80$ MeV are summarized in
Tab. \ref{tab:starting_values}.

 \begin{table}[ht]
\begin{center}
{\renewcommand{\arraystretch}{1.3}
\begin{tabular}{| c | c | c | c |}
 \hline
 $N_f$ &  $a_{10} \, [\mathrm{MeV}^2]$ & $a_{20}$ & $a_{01}$ \\
 \hline
2 & $706.31^2$ & $21.16$ & \\
2+1 & $515.70^2$ & $37.45$ & $47.68$ \\
 \hline 
\end{tabular}}
\end{center}
\caption{FRG input parameters for the chiral ultraviolet potential,
  Eq. \eqref{eq:7}, for $N_f=2$ and $N_f=2+1$ quark
  flavors. \label{tab:starting_values}}
 \end{table}
 
\section{Numerical Solution Details}
\label{sec:numer-solut-deta}

In this work, one of the numerical challenges consists of solving for
multiple gap equations \eqref{eq:gapeq_reformulated} simultaneously,
whereas for each trial point a full FRG flow equation \eqref{eq:frg}
in the (pseudo)scalar sector has to be solved. In order to retain
reasonable computation times, we solve the vector meson gap equations
by computing data points on a discrete set of vector meson condensates
and interpolating key quantities like the number densities
\eqref{eq:number_densities}. This allows us to use the same basic set
of points for all coupling strengths. The same process is used for the
electron chemical potential to satisfy charge neutrality, see
\eqref{eq:14}. Instead of keeping the interpolated values for the
equation of state, we then calculate new data points with the
appropriate shifts in the chemical potentials inserted. This way, we
have a method to check to what extent the gap equations and charge
neutrality condition are actually fulfilled and to thereby gauge the
quality of the interpolation.

The solution of the FRG flow equation \eqref{eq:frg} for the
(pseudo)scalars is performed numerically by discretizing field space
\cite{Bohr2001,Adams1995}. This leads to a set of coupled ordinary
differential equations. Considering recent numerical advances in the
field, see \cite{Grossi:2019urj}, for $N_f=2$ flavors we employ an
upwind finite difference scheme for the determination of field
derivatives while in parallel solving the flow for the potential and
its derivative. For $N_f=2+1$ flavors, the two-dimensional grid of the
two scalar field variables is linked via cubic splines as outlined in
\cite{Mitter:2013fxa} and also applied in preceding works
\cite{Otto:2019zjy}. While we found the latter method to lead to small
numerical inaccuracies for large vector couplings $g_v$ at chemical
potentials higher than the onset of strangeness around
$\mu \approx 430$ MeV, the close agreement with the vastly different
method for $N_f=2$ flavors at smaller chemical potentials gives us
confidence in the general validity of the employed approach.

\bibliography{tovfrg}

\end{document}